\documentclass[aps,prb,twocolumn,showpacs,preprintnumbers,amsmath,amssymb]{revtex4}%
\usepackage{graphicx}% Include figure files
\usepackage{dcolumn}% Align table columns on decimal point
\usepackage{bm}% bold math
\usepackage{color}

\begin{document}

%\preprint{PREPRINT (\today)}

\title{Evidence for Kosterlitz-Thouless and 3D-$xy$ critical behavior in Bi$_{2}$Sr$_{2}$CaCu$_{2}$O$_{8+\delta}$}

\author{S. Weyeneth}
\email{wstephen@physik.uzh.ch}
\affiliation{Physik-Institut der Universit\"{a}t Z\"{u}rich, Winterthurerstrasse 190, CH-8057 Z\"{u}rich, Switzerland.}

\author{T. Schneider}
\affiliation{Physik-Institut der Universit\"{a}t Z\"{u}rich, Winterthurerstrasse 190, CH-8057 Z\"{u}rich, Switzerland.}

\author{E. Giannini}
\affiliation{DPMC, University of Geneva, 24 Quai Ernest-Ansermet, CH-1211 Geneva 4, Switzerland. }

%%%%%%%%%%%%%%%%
\begin{abstract}
%%%%%%%%%%%%%%%%
We present reversible magnetization data of a high quality Bi$_{2}$Sr$_{2}$CaCu$_{2}$O$_{8+\delta}$ single crystal and explore the occurrence of 3D-$xy$ critical behavior close to the bulk transition temperature $T_{\rm c}$ and of Kosterlitz-Thouless (KT) behavior. Below and above the presumed Kosterlitz-Thouless transition temperature $T_{\rm KT}$ we observe the characteristic 2D-$xy$ behavior: a downward shift of the crossing point phenomenon towards $T_{\rm KT}$ as the field is decreased and sufficiently below $T_{\rm KT}$ the characteristic 2D-$xy$ relationship between the magnetization an the in-plane magnetic penetration depth $\lambda_{ab}$. In contrast, the measured temperature dependence of the superfluid density does not exhibit the characteristic KT-behavior around the presumed $T_{\rm KT}$. The absence of this feature is traced back to the 2D- to 3D-$xy$ crossover setting in around and above $T_{\rm KT}$. Invoking the Maxwell relation, the anomalous field dependence of the specific heat peak is also traced back to the intermediate 2D-$xy$ behavior. However, close to $T_{\rm c}$ we observe consistency with 3D-$xy$ critical behavior, in agreement with measurements of $\lambda_{ab}$.
\end{abstract}


\maketitle
%%%%%%%%%%%%%%%%
\section{Introduction}
%%%%%%%%%%%%%%%%
The study of thermal fluctuations received a considerable impetus from the discovery of the cuprate superconductors.\cite{blatter,book,tsben,maxwts, Weyeneth1, Weyeneth2} It was realized that in these materials the critical regime where thermal fluctuations dominate can be attained and that some of them are in addition quasi two dimensional (2D).\cite{gamma} Furthermore, the systematics of the superconducting properties uncovered that the anisotropy is further enhanced with underdoping.\cite{book,gamma} In this quasi 2D limit one expects the thermodynamic properties to be close to those of a two-dimensional superconductor, or more precisely of a stack of decoupled two-dimensional superconducting sheets. Although approximate treatments have been invoked to describe the thermodynamic properties of such materials, the essential ingredient, the Kosterlitz-Thouless behavior of the associated zero field transition,\cite{kosterlitz} has mostly not been taken into account.\cite{bula,mosq,vidal} Only recently, this behavior was incorporated by combining Kosterlitz-Thouless (KT) renormalization-group flows and explicit computations for plasmas.\cite{oganesyan} On this basis the field and temperature dependence of the magnetization density, $m\left(H_{c},T\right) $, for temperatures $T$ near to and below the Kosterlitz-Thouless (KT) transition temperature $T_{\rm KT}$ was determined for magnetic fields $H_{c}$ applied perpendicular to the superconducting sheet. These results are interesting on three immediate fronts. First, the resulting trends in the magnetization appear to emerge from the recent Bi$_{2}$Sr$_{2}$CaCu$_{2}$O$_{8+\delta }$ data of Lu Li \textit{et al}. for underdoped and optimally doped samples.\cite{luli} Second, by contrast evidence for smeared 3D-$xy$ behavior stems from the measured temperature dependence of the in-plane magnetic penetration depth $\lambda_{ab}$.\cite{jacobs,lee,osborn,tscastro} Third, the magnetic field dependence of the specific heat peak exhibits, opposite to the generic behavior,\cite{maxwts, Weyeneth2} a shift to higher temperatures with increasing field strength.\cite{junod}
%%%%%%%%%%%%%%%%

In this study we present reversible magnetization data of a Bi$_{2}$Sr$_{2}$CaCu$_{2}$O$_{8+\delta }$ single crystal and explore the evidence for intermediate KT- (2D-$xy$) and 3D-$xy$ critical behavior. Below $T=89.5$ K $\simeq T_{\rm KT}$ we observe consistency with the KT-behavior in terms of the characteristic $m\propto \ln (H_{c})$ dependence at fixed temperature. Furthermore, invoking the Maxwell relation $\partial^{2}M/\partial T^{2}\vert _{H_{c}}=\partial\left(C/T\right)/\partial H_{c}\vert _{T}$ the anomalous field dependence of the specific heat peak is also traced back to 2D-$xy$ behavior. However, close to the bulk transition temperature, $T_{\rm c}\simeq 91.21$ K we observe consistency with 3D-$xy$ critical behavior, consistent with previous measurements of the in-plane magnetic penetration depth $\lambda_{ab}$.\cite{jacobs,osborn,tscastro} In Section II we sketch the theoretical background including the scaling relations for 2D- and 3D-$xy$ critical behavior. Section III is devoted to the experimental details and in Section IV we present the analysis of the data uncovering the evidence for 2D-$xy$ and 3D-$xy$ critical behavior in the respective temperature regimes. We close with a brief summary and some discussion.
%%%%%%%%%%%%%

%%%%%%%%%%%%%%%%
\section{Theoretical background}
%%%%%%%%%%%%%%%%
When thermal fluctuations dominate and the coupling to the charge is negligible a bulk superconductor is expected to exhibit sufficiently close to $T_{\rm c}$ 3D-$xy$ critical behavior. In this case the magnetization per unit volume, $m=M/V$, adopts the scaling form\cite{book,tsben,maxwts, Weyeneth1, Weyeneth2,jhts}
\begin{eqnarray}
\frac{m}{TH_c^{1/2}} &=&-\frac{Q^{\pm}k_{\rm B}\xi _{ab}}{\Phi _{0}^{3/2}\xi _{c}}F^{\pm}(z),\text{ }F^{\pm}(z)=z^{-1/2}\frac{dG^{\pm}}{dz},  \nonumber \\
z &=&x^{-1/2\nu}=\frac{(\xi _{ab0}^{\pm})^{2}|t|^{-2\nu}H_{c}}{\Phi_{0}}. 
\label{eq1}
\end{eqnarray}
In this form $Q^{\pm }$ is a universal constant and $G^{\pm }\left( z\right) $ a universal scaling function of its argument, with $G^{\pm }\left( z=0\right)=1$. In addition $\gamma =\xi _{ab}/\xi _{c}$ denotes the anisotropy, $\xi _{ab}$ the zero-field in-plane correlation length and $H_{c}$ the magnetic field applied along the $c$-axis. In terms of the variable $x$ the scaling form (\ref{eq1}) is similar to Prange's result for Gaussian fluctuations.\cite{prange} Approaching $T_{\rm c}$ the correlation lengths diverges as
\begin{equation}
\xi _{ab,c}=\xi _{ab0,c0}^{\pm}|t|^{-\nu},\text{ }t=T/T_{\rm c}-1,\text{ }\pm=sgn(t).  
\label{eq2}
\end{equation}
Supposing that 3D-$xy$ fluctuations dominate the critical exponents are given by\cite{pelissetto}
\begin{equation}
\nu\simeq0.671\simeq2/3,\text{ }\alpha =2\nu-3\simeq-0.013,  
\label{eq3}
\end{equation}
and there are the universal critical amplitude relations\cite{book,tsben,maxwts,jhts,pelissetto}
\begin{equation}
\frac{\xi _{ab0}^{-}}{\xi _{ab0}^{+}}=\frac{\xi _{c0}^{-}}{\xi _{c0}^{+}}\simeq 2.21,\text{ }\frac{Q^{-}}{Q^{+}}\simeq 11.5,\text{ }\frac{A^{+}}{A^{-}}=1.07,  
\label{eq4}
\end{equation}
and
\begin{eqnarray}
A^{-}\xi _{a0}^{-}\xi _{b0}^{-}\xi _{c0}^{-}&\simeq&A^{-}(\xi_{ab0}^{-})^{2}\xi _{c0}^{-}=\frac{A^{-}(\xi _{ab0}^{-})^{3}}{\gamma}  \nonumber \\
&=&(R^{-})^{3},R^{-}\simeq 0.815,  
\label{eq5}
\end{eqnarray}
where $A^{\pm}$ is the critical amplitude of the specific heat singularity, defined as
\begin{equation}
c=\frac{C}{Vk_{\rm B}}=\frac{A^{\pm}}{\alpha}|t|^{-\alpha}+B,  
\label{eq6}
\end{equation}
where $B$ denotes the background. The anisotropy is then characterized in terms of
\begin{equation}
\gamma =\frac{\xi _{ab}}{\xi _{c}}=\frac{\xi _{ab0}^{\pm }}{\xi _{c0}^{_{\pm}}}.  
\label{eq7}
\end{equation}
Furthermore, in the 3D-$xy$ universality class $T_{\rm c}$, $\xi _{c0}^{-}$ and the critical amplitude of the in-plane magnetic penetration depth $\lambda _{ab0}$ are not independent but related by the universal relation,\cite{book,tsben,maxwts,jhts}
\begin{equation}
k_{\rm B}T_{\rm c}=\frac{\Phi _{0}^{2}}{16\pi ^{3}}\frac{\xi _{c0}^{-}}{\lambda_{ab0}^{2}}=\frac{\Phi _{0}^{2}}{16\pi ^{3}}\frac{\xi _{ab0}^{-}}{\gamma\lambda _{ab0}^{2}}.  
\label{eq8}
\end{equation}
The existence of the magnetization at $T_{\rm c}$, of the magnetic penetration depth below $T_{\rm c}$ and of the magnetic susceptibility above $T_{\rm c}$ imply the following asymptotic forms of the scaling function\cite{book,tsben,maxwts,jhts}
\begin{eqnarray}
Q^{\pm}\left. \frac{1}{\sqrt{z}}\frac{dG^{\pm}}{dz}\right\vert_{z\rightarrow\infty} &=& Q^{\pm}c_{\infty}^{\pm},  \nonumber \\
Q^{-}\left. \frac{dG^{-}}{dz}\right\vert_{z\rightarrow0} &=& Q^{-}c_{0}^{-}(\ln z+c_{1}) ,  \nonumber \\
Q^{+}\left. \frac{1}{z}\frac{dG^{+}}{dz}\right\vert_{z\rightarrow0} &=& Q^{+}c_{0}^{+},  
\label{eq9}
\end{eqnarray}
with the universal coefficients
\begin{equation}
Q^{-}c_{0}^{-}\simeq-0.7,\text{ }Q^{+}c_{0}^{+}\simeq0.9,\text{ }Q^{\pm}c_{\infty}^{\pm}\simeq0.5,\text{ }c_{1}\simeq1.76.  
\label{eq10}
\end{equation}
%%%%%%%%%%%%%%%%

Noting that Bi$_{2}$Sr$_{2}$CaCu$_{2}$O$_{8+\delta}$ is highly anisotropic $\left(\gamma >>1\right) $,\cite{watauchi} the system is expected to exhibit away from $T_{\rm c}$ 2D-$xy$ behavior. A characteristic property of 2D-superconductors emerges from the magnetic field dependence of the magnetization. Sufficiently below the Kosterlitz-Thouless transition temperature $T_{\rm KT}$ the magnetization is given by\cite{oganesyan}
\begin{eqnarray}
m &=&-\frac{\pi \rho_{\rm s}\left(T\right) }{2d\Phi _{0}}\left(1-\frac{k_{\rm B}T}{\pi\rho_{\rm s}\left(T\right)}\right)\ln\left(\frac{\Phi _{0}}{4\pi H_{c}a_{0}^{2}\gamma _{3}}\right)  \nonumber \\
&=&-\left(\frac{\Phi_{0}}{32\pi^{2}\lambda_{ab}^{2}\left(T\right)}-\frac{k_{\rm B}T}{d\Phi_{0}}\right)\ln\left(\frac{\Phi _{0}}{4\pi H_{c}a_{0}^{2}\gamma _{3}}\right) ,  
\label{eq11}
\end{eqnarray}
where $\rho_{\rm s}$ is the 2D superfluid density, related to the in-plane magnetic penetration depth $\lambda_{ab} $ via
\begin{equation}
\rho_{\rm s}(T) =\frac{d\Phi _{0}^{2}}{16\pi^{3}\lambda_{ab} ^{2}(T)}.  
\label{eq12}
\end{equation}
Here $d$ is the thickness of the independent superconducting sheets, $\gamma _{3}$ is a parameter that vanishes as $T$ approaches $T_{\rm KT}$, and $a_{0}$ the microscopic short-distance cutoff length. Moreover, $\rho_{\rm s}\left(T_{\rm KT}\right) $ and $T_{\rm KT}$ are related by
\begin{equation}
\rho_{\rm s}\left(T_{\rm KT}\right) =\frac{2}{\pi }k_{\rm B}T_{\rm KT},  
\label{eq13}
\end{equation}
while $\rho_{\rm s}\left(T\right)=0$ above $T_{\rm KT}$. Below this universal Nelson-Kosterlitz jump $1/\lambda_{ab}^{2}\left(T\right)$ increases as\cite{nelson,ambek}
\begin{equation}
\lambda_{ab} ^{2}\left(T_{\rm KT}\right)/\lambda_{ab} ^{2}\left(T\right)=\frac{T}{T_{\rm KT}}\left(1+\widehat{b}\left(T_{\rm KT}-T\right)^{1/2}\right).
\label{eq14}
\end{equation}
Note that Eq. (\ref{eq11}) gives a simple relation between the superfluid density and the derivative of the magnetization, namely
\begin{equation}
\frac{dm}{d\ln(H_{c})}=g\left(T\right)=\frac{\Phi_{0}}{32\pi ^{2}\lambda_{ab}^{2}\left(T\right)}-\frac{k_{\rm B}T}{d\Phi_{0}}. 
\label{eq15}
\end{equation}
Furthermore, at criticality $m$ depends on $H$ in terms of\cite{oganesyan}
\begin{equation}
m=-\frac{k_{\rm B}T_{\rm KT}}{d\Phi_{0}}\ln\left(\gamma_{1}\ln\frac{\Phi_{0}}{4\pi H_{c}a_{0}^{2}\gamma_{2}}\right),  
\label{eq16}
\end{equation}
where $\gamma_{1}$ and $\gamma_{2}$ are constants. Correspondingly, plots of $M$ \textit{vs}. $T$ at different fields should then exhibit a systematic drift of the \textquotedblleft crossing phenomenon\textquotedblright. Above $T_{\rm KT}$ and for asymptotically small fields the magnetization is given by\cite{oganesyan,halperin,ts07}
\begin{equation}
m\simeq-\frac{k_{\rm B}T}{2d\Phi_{0}^{2}}\xi_{ab}^{2}H_{c},\text{ }\xi_{ab}=\xi_{ab0}\exp\left(\frac{\widetilde{b}}{\left(T/T_{\rm KT}-1\right)^{1/2}}\right), 
\label{eq17}
\end{equation}
where $\xi_{ab}$ is the Kosterlitz-Thouless correlation length.\cite{kosterlitz} The parameters $\widetilde{b}$ and $\widehat{b}$, determining the temperature dependence of the magnetic penetration depth below the jump (Eq. (\ref{eq14})), are related by\cite{ambek}
\begin{equation}
\widetilde{b}\widehat{b}=\pi/\left(2T_{\rm KT}^{1/2}\right)\simeq0.17\text{ K}^{-1/2}.
\label{eq18}
\end{equation}
%%%%%%%%%%%%%%%%

%%%%%%%%%%%%%%%%
\begin{figure}[t!]
\includegraphics[width=1\linewidth]{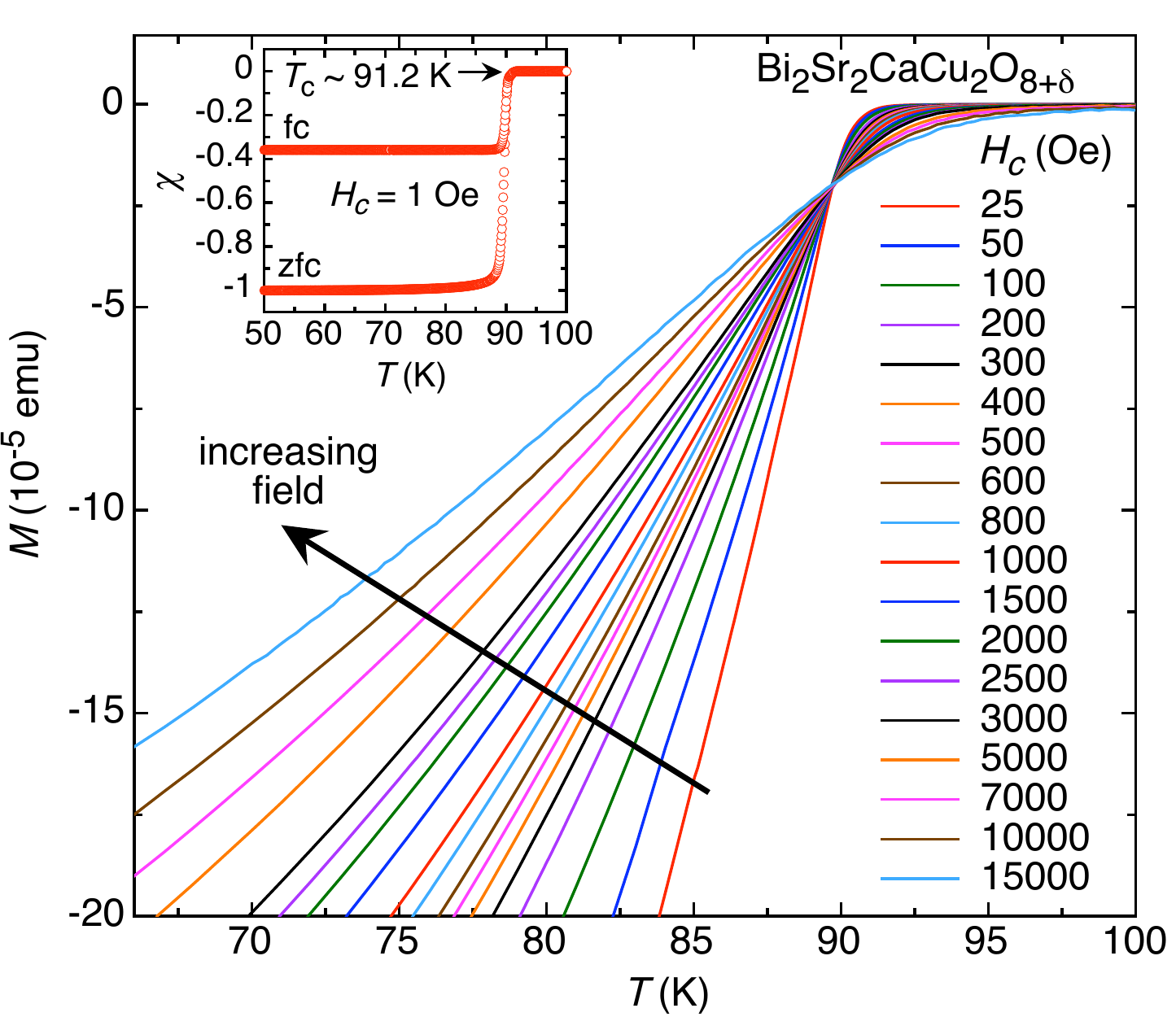}
\vspace{0cm}
\caption{(color online) Temperature dependence of the measured reversible magnetic moment of the studied Bi$_{2}$Sr$_{2}$CaCu$_{2}$O$_{8+\delta }$ single crystal at various magnetic fields applied along the $c$-axis. The inset depicts susceptibility measurements in 1 Oe in the ZFC and FC mode. The sharp onset of superconductivity points to a transition temperature close to $T_{\rm c}\simeq 91.2$ K.}
\label{fig1}
\end{figure}
%%%%%%%%%%%%%%%%

%%%%%%%%%%%%%%%%
\section{Experimental Details}
%%%%%%%%%%%%%%%%
Single crystals of Bi$_2$Sr$_2$CaCu$_2$O$_{8+\delta}$ were grown with the Floating Zone (FZ) method, from direct crystallization from the melt (no solvent used). The feed rod was obtained from a commercial Bi$_{2}$Sr$_{2}$CaCu$_{2}$O$_{8+x}$ powder after pressing and sintering to a shape of 7 cm in length and about 7 mm in diameter. The seed rod was made with a previously crystallized rod. After a first fast FZ melting at a rate of 24 mm/h in Ar, the crystal growth was performed at a slow rate of 0.2 mm/h in a $7$\% O$_{2}-93$\% Ar atmosphere, while both rods were counter-rotating at 18 rpm. The FZ growth was performed in a commercial two-mirror vertical furnace (from Cyberstar), equipped with two 1000 W halogen lamps. The growth conditions at the flat zone interface were kept stable for several days. A review of the growth technique can be found elsewhere.\cite{revco} The as-grown crystals were easily cleaved from the crystallized boule and annealed at $T=500^{\circ }$C for $50$h in $0.1$\% O$_{2}-99.9$\% Ar, in order to tune and homogenize the oxygen content corresponding to the optimal doping level. Crystals with typical size of $1-5$ mm and thickness of $0.05-0.1$ mm could be extracted. The good crystalline quality of the samples was checked by X-ray diffraction and magnetic susceptibility measurements.
%%%%%%%%%%%%%%%%

%%%%%%%%%%%%%%%%
\begin{figure}[b!]
\includegraphics[width=1\linewidth]{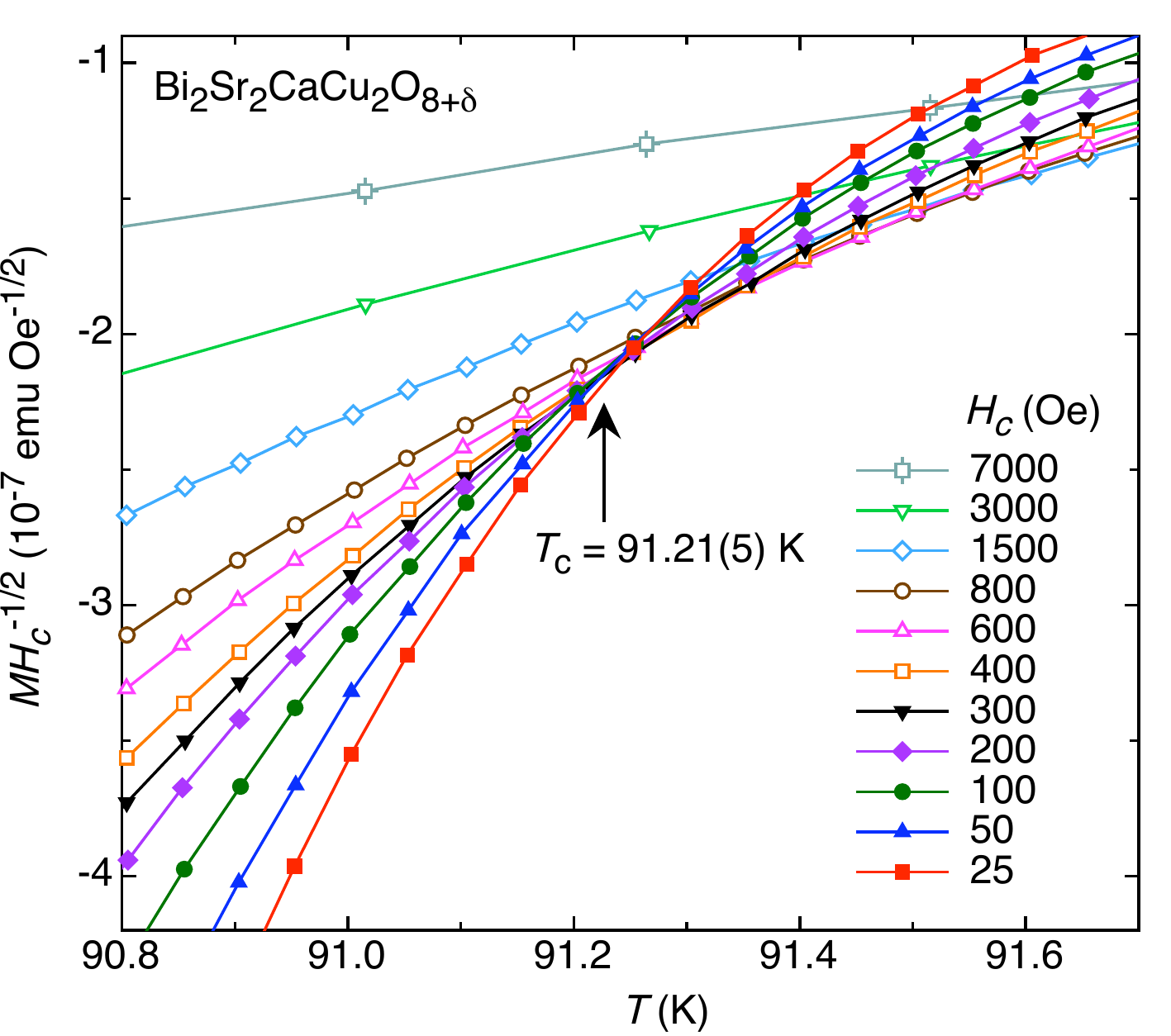}
\vspace{0cm}
\caption{(color online) $M/H_{c}^{1/2}$ \textit{vs}. $T$ yielding the estimate $T_{\rm c}\simeq 91.21$ K in terms of the crossing point at $M/H_{c}^{1/2}\simeq-2.13\times 10^{-7}$ emuOe$^{-1/2}$.}
\label{fig2}
\end{figure}
%%%%%%%%%%%%%%%%

The Bi$_{2}$Sr$_{2}$CaCu$_{2}$O$_{8+\delta}$ sample used in this work, a $V\simeq 4.6\times 10^{-5}$ cm$^{3}$ single crystal, was chosen by its sharp low-field Meissner transition from several high quality single crystals. The magnetization was measured in a Quantum Design DC-SQUID magnetometer MPMS XL with an installed Reciproating Sample Option. The inset in Fig. \ref{fig1} shows the measured susceptibility at $H_{c}=1$ Oe applied along the $c$-axis. It reveals a rather sharp transition at $T_{\rm c}\simeq 91.2$ K and a well saturated Meissner state, pointing to excellent quality. The volume of the sample was estimated by susceptibility measurements below $T_{\rm c}$ in the Meissner state with a magnetic field applied along the $ab$- plane to minimize demagnetization effects. The extracted volume of $V\simeq 4.6\times10^{-5}$ cm$^{3}$ compares well with that estimated with an optical microscope. Fig. \ref{fig1} summarizes the measured temperature dependence of the magnetic moment at fields ranging from $25$ Oe to $15000$ Oe applied along the $c$-axis. After applying the magnetic field, well below $T_{\rm c}$ it was kept constant and the magnetic moment of the single crystal was measured at a stabilized temperature by moving the sample with a frequency of $0.5$ Hz through a set of detection coils. The reversible superconducting diamagnetic magnetization, $M=mV$, was then obtained by comparing field cooled (FC) and zero-field cooled (ZFC) data. Due to a substantial pinning contribution at low magnetic field we omitted data below 25 Oe. A temperature dependent normal state paramagnetic background was subtracted.
%%%%%%%%%%%%%%%%

%%%%%%%%%%%%%%%%
\section{Data analysis}
%%%%%%%%%%%%%%%%
We are now prepared to analyze the magnetization data. To estimate the bulk $T_{\rm c}$ we invoke Eqs. (\ref{eq1}), (\ref{eq9}) and (\ref{eq10}), revealing that the plot $m/H_{c}^{1/2}$ \textit{vs}. $T$ should exhibit a crossing point at $T_{\rm c}$. Here $m/(TH_{c}^{1/2})$ adopts with Eq. (\ref{eq7}) the value $m/(T_{\rm c}H_{c}^{1/2})=-0.5k_{\rm B}\gamma \Phi _{0}^{-3/2}$. According to Fig. \ref{fig2}, showing $M/H_{c}^{1/2}$ \textit{vs}. $T$ there is a crossing point at $T_{\rm c}\simeq 91.21$ K where $M/H_{c}^{1/2}\simeq -2.13\times 10^{-7}$ emuOe$^{-1/2}$. With $V\simeq 4.6\times 10^{-5}$ cm$^{3}$, where $m=M/V$, it yields for the anisotropy (Eq. (\ref{eq7})) the estimate 
\begin{equation}
\gamma =\xi_{ab0}^{-}/\xi_{c0}^{-}\simeq 69,
\label{eq19}
\end{equation}
compared to $\gamma\simeq 133$ for an underdoped sample with $T_{\rm c}\simeq 84.2$ K and in reasonable agreement with earlier estimates for optimally doped samples.\cite{watauchi,piriou} Given this rather large anisotropy the 2D- to 3D-$xy$ crossover is expected to occur rather close to the bulk $T_{\rm c}$.
%%%%%%%%%%%%%%%% 

\begin{figure}[b!]
\includegraphics[width=1\linewidth]{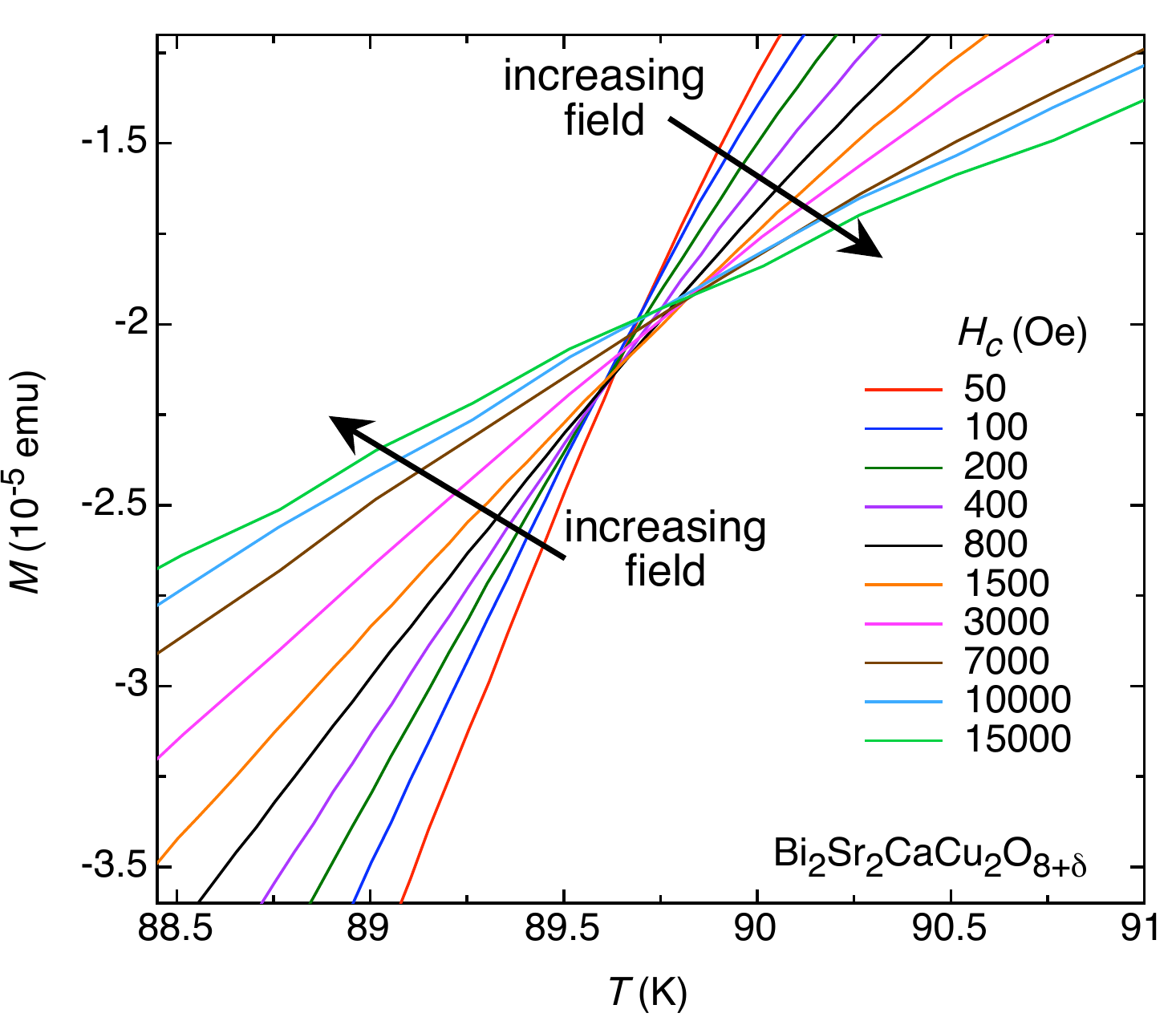}
\vspace{0cm}
\caption{(color online) $M$ \textit{vs}. $T$ at various magnetic fields exhibiting a crossing phenomenon shifting towards $T_{\rm KT}\simeq 89.5$ K from above as the field is decreased.}
\label{fig3}
\end{figure}

A characteristic feature of a 2D superconductor is the crossing phenomenon occurring at fixed magnetic fields and above $T_{\rm KT}$ in the plot $M$ \textit{vs}. $T$.\cite{book,oganesyan} A glance at Fig. \ref{fig3} reveals that this phenomenon is well confirmed above $T=89.5$ K $\simeq T_{\rm KT}$. Indeed, there is as predicted a downward shift of the \textquotedblleft crossing point\textquotedblright\ towards $T_{\rm KT}$ from above as the field is decreased.\cite{oganesyan} The same behavior was also observed in the highly anisotropic Tl-1223, Bi-2201 and underdoped La$_{2-x}$Sr$_{x}$CuO$_{4}$ single crystals.\cite{book,gtris,gtris2,iwasaki}
%%%%%%%%%%%%%%%%

%%%%%%%%%%%%%%%%
%%%%%%%%%%%%%%%%

%%%%%%%%%%%%%%%%
\begin{figure}[t!]
\includegraphics[width=1\linewidth]{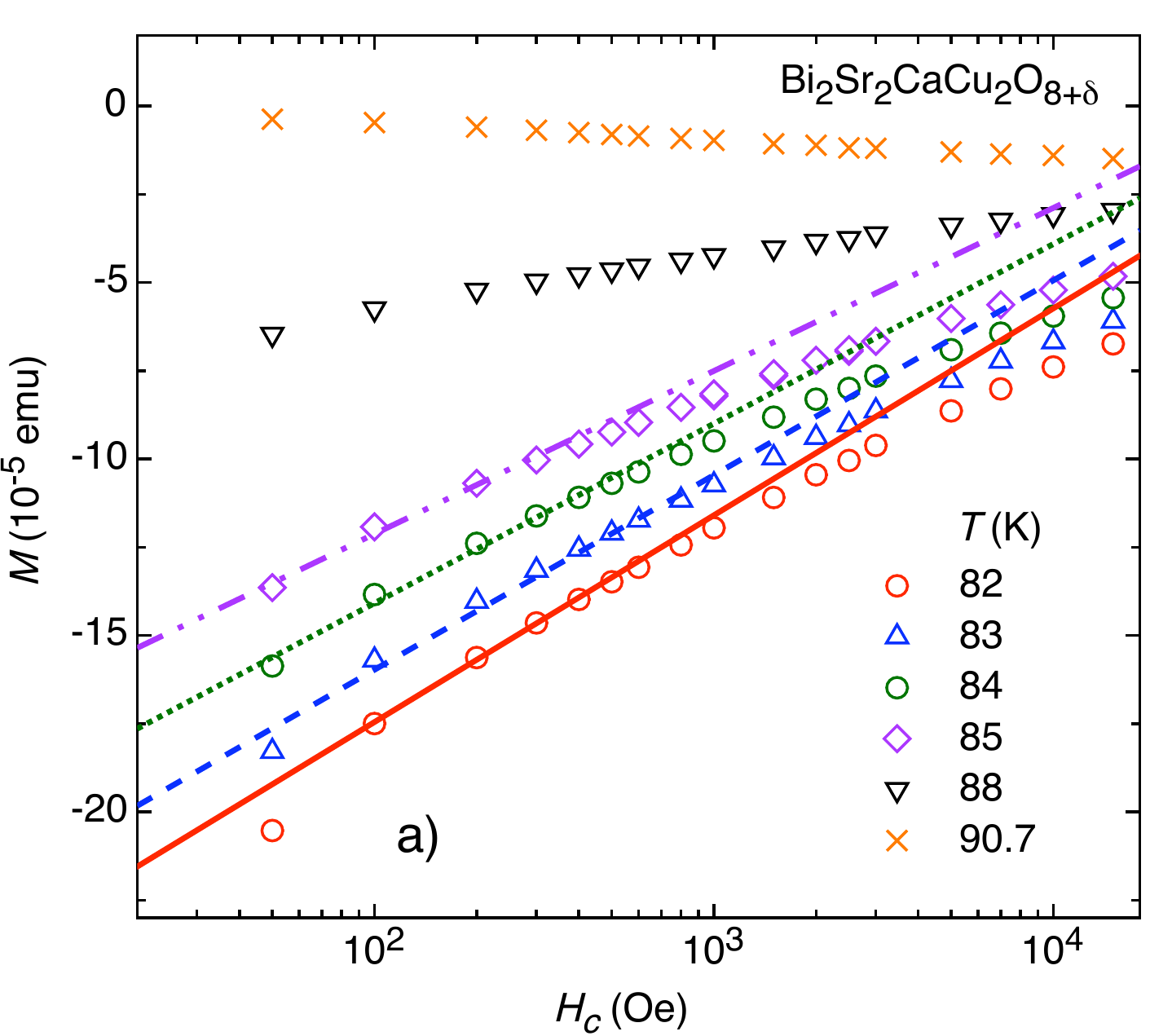}
\includegraphics[width=1\linewidth]{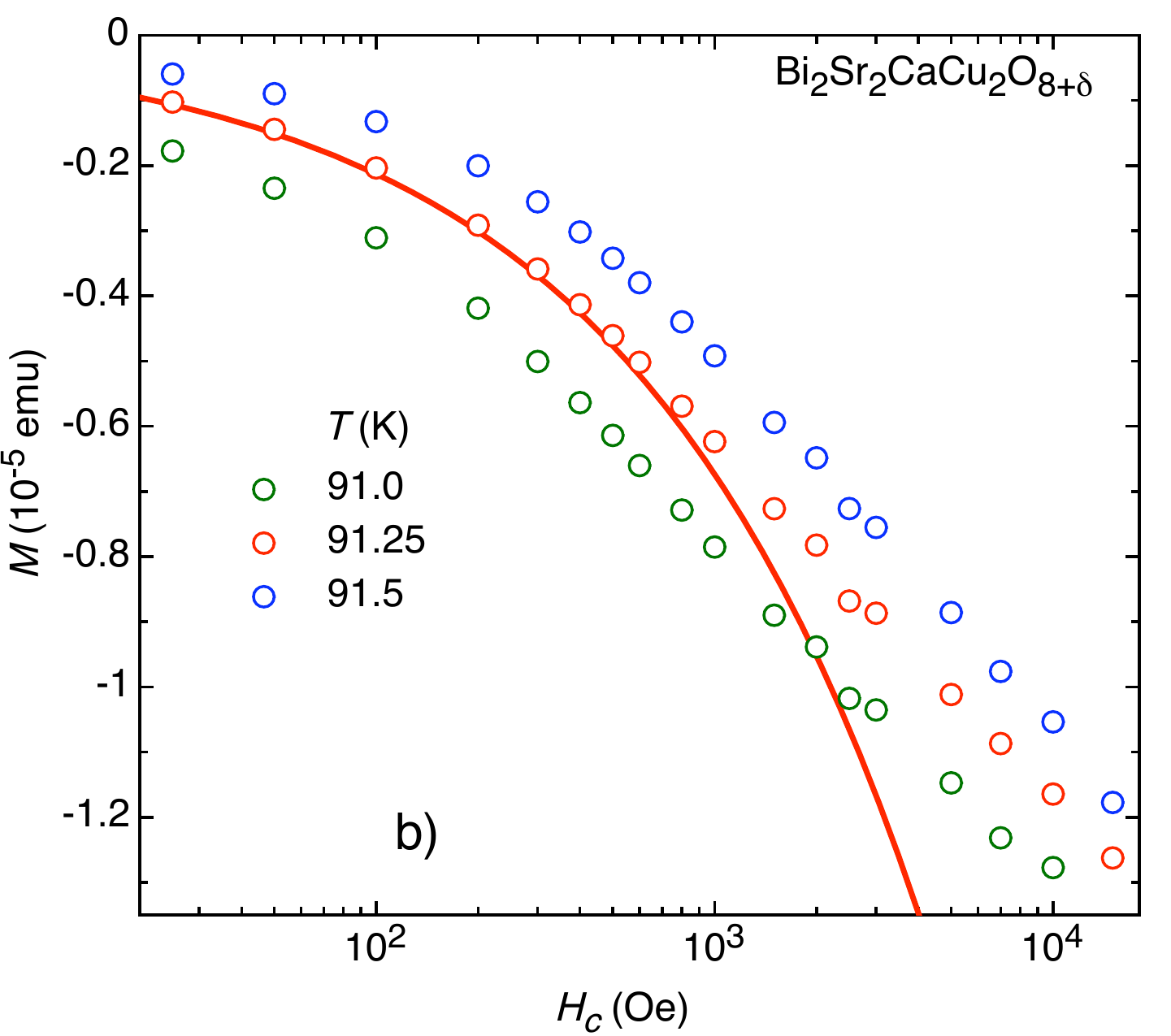}
\vspace{0cm}
\caption{(color online) $M$ \textit{vs}. $H_{c}$ at various fixed temperatures. a) from $T=82$ to $90.7$ K. The straight lines indicate the characteristic 2D relation between the superfluid density and the derivative of the magnetization with respect to the logarithm of the field (Eq. (\ref{eq15})), valid in the low field regime as long as pinning contributions are negligible. b) $M$ \textit{vs}. $H_{c}$ at $T=91$, $91.25$ K and $91.5$ K. The solid line is $M=-2.13\times 10^{-7}\cdot H_{c}^{1/2}$ emu indicating the characteristic 3D-$xy$ behavior at $T_{\rm c}$ in terms of Eq. (\ref{eq20}).}
\label{fig4}
\end{figure}
%%%%%%%%%%%%%%%%

To substantiate 2D-$xy$ behavior further we invoke Eq. (\ref{eq11}) in terms of the plot $M$ \textit{vs}. $H_c$ shown in Fig. \ref{fig4}. The solid lines in Fig. \ref{fig4}a indicate that sufficiently below $T_{\rm KT}\simeq 89.5$ K the 2D relation (\ref{eq15}) between the superfluid density and the derivative of the magnetization with respect to the logarithm of the field is for small fields well obeyed, so in this temperature regime the system indeed behaves as a stack of essentially decoupled two dimensional Kosterlitz-Thouless films with $T_{\rm KT}$ $\simeq 89.5$ K. However, close to the bulk $T_{\rm c}\simeq 91.21$ K one expects 3D-$xy$ critical behavior. In this case and in the magnetic field range considered here the limit $z\rightarrow \infty $ is then approached. Here Eqs. (\ref{eq1}), (\ref{eq9}) and (\ref{eq10}) imply the limiting behavior
\begin{equation}
m=-\frac{Q^{+}c_{\infty }^{+}k_{\rm B}\xi _{ab}T}{\Phi _{0}^{3/2}\xi _{c}}H_{c}^{1/2}\simeq -\frac{0.5k_{\rm B}\xi _{ab0}^{+}T}{2\Phi _{0}^{3/2}\xi_{c0}^{+}}H_{c}^{1/2}.
\label{eq20}
\end{equation}
%%%%%%%%%%%%%%%%

Fig. \ref{fig4}b, depicting $M$ \textit{vs}. $H_{c}$ close to $T_{\rm c}\simeq 91.21$ K at $T=91$, $91.25$ and $91.5$ K, shows that this expectation is well confirmed. According to this, approaching $T_{\rm c}$ the system undergoes a 2D-to 3D-$xy$ crossover. It implies that the characteristic 2D-$xy$ critical behavior, including the jump of the superfluid density at $T_{\rm KT}$ (Eqs. (\ref{eq13}) and (\ref{eq14})), are removed.
%%%%%%%%%%%%%%%%

%%%%%%%%%%%%%%%%
\begin{figure}[t!]
\includegraphics[width=1\linewidth]{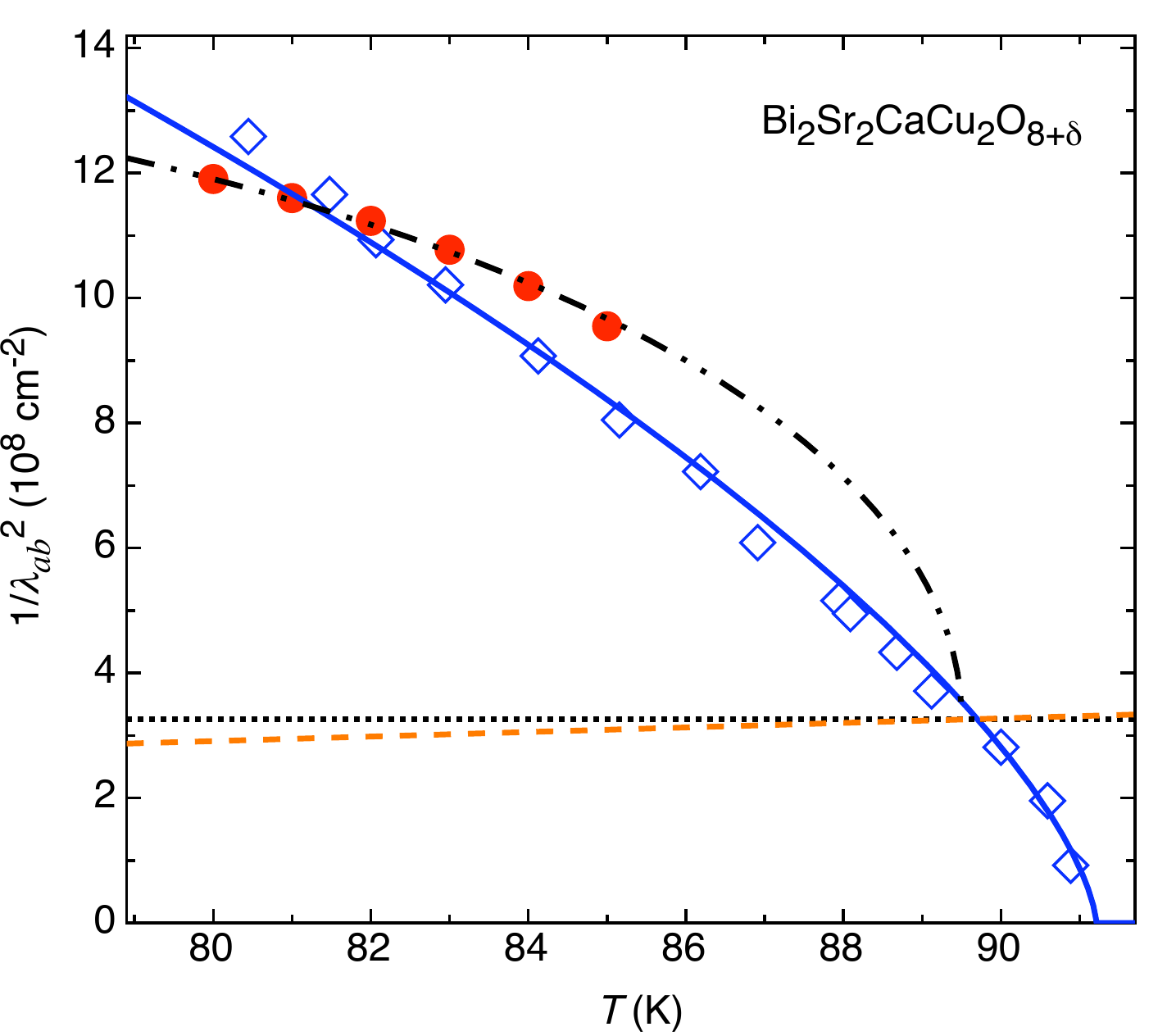}
\vspace{0cm}
\caption{(color online) $1/\lambda _{ab}^{2}\left( T\right) $ $vs$. $T$. $\bullet$ :derived from the magnetization with the aid of Eq. (\ref{eq11}) rewritten as $M(H_{c},T)=f(T)+g(T)\cdot\ln(H_{c})$  and $d=30$ \AA ; $\blacksquare$: Experimental data derived from Lee \textit{et al}. with $\lambda _{ab}\left( T=0\right) =1350$\AA.\cite{lee} The solid line is $1/\lambda _{ab}^{2}\left( T\right) =1/\lambda_{ab0}^{2}\left( 1-T/T_{\rm c}\right) ^{2/3}$ with $1/\lambda _{ab0}^{2}=5\times10^{9}$ cm$^{-2}$ and $T_{\rm c}=91.21$ K indicating 3D-$xy$ critical behavior. The dashed line is the KT-line $1/\lambda _{ab}^{2}\left( T\right)=3.64\times 10^{6}\cdot T$ cm$^{-2}$(Eq. (\ref{eq13})), the dotted one $1/\lambda _{ab}^{2}\left( T_{\rm KT}\right) =3.26\times 10^{8}$ cm$^{-2}$ with $T_{\rm KT}=89.5$ K and the dash-dot-dot one is Eq. (\ref{eq14}) with $\widehat{b}\simeq1$ K$^{-1/2}$.}
\label{fig5}
\end{figure}
%%%%%%%%%%%%%%%%

To clarify this point we invoke Eq. (\ref{eq11}) to determine the temperature dependence of $1/\lambda _{ab}^{2}$ sufficiently below $T_{\rm KT}$. Setting $M(H_{c},T)=f(T)+g(T)\cdot\ln(H_{c})$, $g(T)$ is then given by Eq. (\ref{eq15}) and follows from plots as shown in Fig. \ref{fig4}a in terms of the slope of the straight lines. For $d=30$ \AA\ we obtain for $1/\lambda _{ab}^{2}\left(T\right) $ the data points shown in Fig. \ref{fig5}. For comparison we included the $ab$-plane microwave surface impedance data of Lee \textit{et al}. for a high-quality Bi$_{2}$Sr$_{2}$CaCu$_{2}$O$_{8}$ single crystal, plotted in terms of the superfluid density, assuming $\lambda _{ab}(0)=1350$ \AA.\cite{lee} Sufficiently below $T_{\rm KT}$ we observe reasonable agreement with the 2D-$xy$ prediction ($\bullet $) and the measured $1/\lambda _{ab}^{2}\left( T\right)$ $\left( \blacksquare \right) $. Accordingly, in this regime the expulsion of vortices from the KT phase dominate. Here we also observe consistency with the characteristic KT-behavior indicted by the dash-dot-dot line (Eq. (\ref{eq14})). However, around $T_{\rm KT}$, indicated by the horizontal and KT-line, the measured $1/\lambda _{ab}^{2}\left( T\right) $ \ does not exhibit any evidence for the characteristic jump from $1/\lambda _{ab}^{2}\left(T_{\rm KT}\right) $ to zero for $T>T_{\rm KT}$. In contrast agreement with the leading 3D-$xy$ critical behavior is observed (solid line), revealing the removal of the characteristic jump in $1/\lambda _{ab}^{2}\left( T\right) $ at $T_{\rm KT}$ due to the 2D- to 3D-$xy$ crossover. To explore the effect of the 3D-$xy$ fluctuations further, we use the critical amplitude, $1/\lambda_{ab0}^{2}=5\times 10^{9}$ cm$^{-2}$ and $T_{\rm c}=91.21$ K to obtain from the universal relation (\ref{eq8}) the estimate
\begin{equation}
\xi _{c0}^{-}\simeq 2.9\text{ \AA },
\label{eq21}
\end{equation}
for the amplitude of the $c$-axis correlation length. The occurrence of 3D-$xy$ critical behavior then requires that $\xi _{c}\left( T\right) =\xi_{c0}^{-}\left\vert t\right\vert ^{-2/3}$ exceeds $d$ considerably, yielding with $d=30$ \AA\ and $T_{\rm c}=91.21$ K for the onset of 3D fluctuations the lower bound $T>90.24$ K. Furthermore, around and above $T=91$ K is the 3D-$xy$ critical regime is attained (see Fig. \ref{fig4}b). So the occurrence of KT- behavior in $1/\lambda _{ab}^{2}\left(T\right) $ is restricted to temperatures below $T_{\rm KT}\simeq 89.5$ K.
%%%%%%%%%%%%%%%%

On the other hand, the small amplitude of the $c$-axis correlation length leads with $\gamma =69$ (Eq. (\ref{eq19})) to a rather large amplitude of the in-plane correlation length, $\xi _{ab0}^{-}=\gamma \xi _{c0}^{-}\simeq200$ \AA\ and with the universal relation (\ref{eq5}) to a very small amplitude of the specific heat singularity, 
\begin{equation}
A^{-}=\frac{0.815^{3}}{\gamma ^{2}\left( \xi _{c0}^{-}\right) ^{3}}\simeq4.7\times 10^{-6}\text{\AA }^{-3},
\label{eq22}
\end{equation}
in comparison with $A^{-}\simeq 6.8\times 10^{-4}$\AA $^{-3}$ in optimally doped YBa$_{2}$Cu$_{3}$O$_{7-\delta }$ and $1.7\times 10^{-3}$\AA $^{-3}$ in $^{4}$He.\cite{book} This small critical amplitude renders it difficult to observe 3D-$xy$ critical behavior in the specific heat of Bi$_{2}$Sr$_{2}$CaCu$_{2}$O$_{8+\delta }$.\cite{junod}
%%%%%%%%%%%%%%%%

%%%%%%%%%%%%%%%%
\begin{figure}[t!]
\includegraphics[width=1\linewidth]{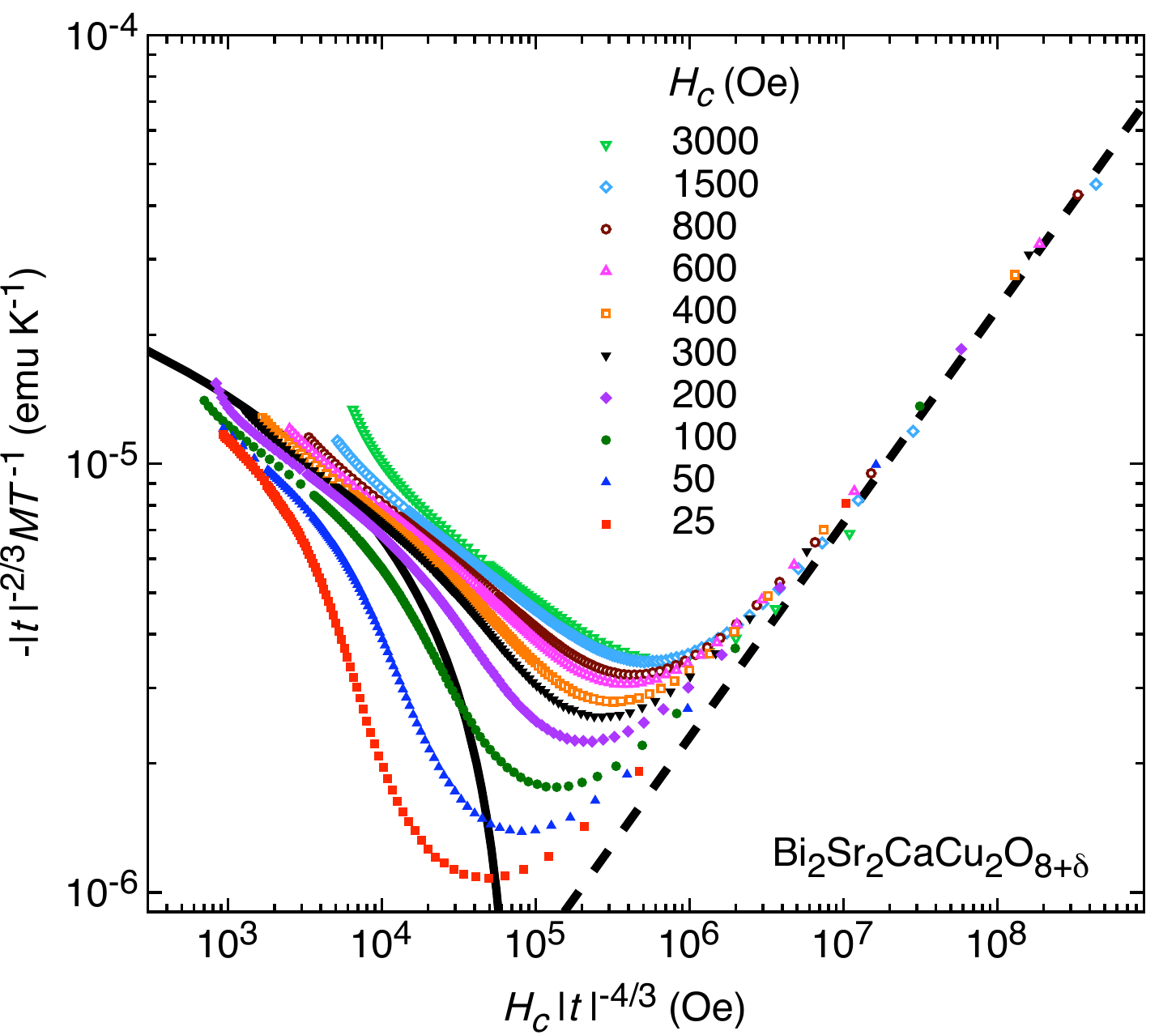}
\vspace{0cm}
\caption{(color online) $-\left\vert t\right\vert ^{-2/3}M/T$ \textit{vs}. $H_{c}\left\vert t\right\vert ^{-4/3}$ for $T<T_{\rm c}$ at various applied magnetic fields. The solid line is Eq. (\ref{eq25}) and the dashed one Eq. (\ref{eq26}).}
\label{fig6}
\end{figure}
%%%%%%%%%%%%%%%%

Considering Eqs. (\ref{eq1}), (\ref{eq9}) and (\ref{eq10}) consistency with 3D-$xy$ critical behavior also requires that the data scales below $T_{\rm c}$ as
\begin{eqnarray}
\left\vert t\right\vert ^{-2/3}\frac{m}{T}&=&-\frac{Q^{-}c_{0}^{-}k_{\rm B}}{\Phi_{0}\xi _{c0}^{-}}\cdot \label{eq23}\\ &\textrm{}& \left( \ln \left( \frac{\left( \xi _{ab0}^{-}\right) ^{2}}{\Phi _{0}}H_{c}\left\vert t\right\vert ^{-4/3}\right) +c_{1}\right) ,\text{ }z\rightarrow 0, \nonumber
\end{eqnarray}
and
\begin{equation}
\left\vert t\right\vert ^{-2/3}\frac{m}{T}=-\frac{Q^{-}c_{\infty}^{-}k_{\rm B}\xi _{ab0}^{-}}{\Phi _{0}^{3/2}\xi _{c0}^{-}}\left(H_{c}\left\vert t\right\vert ^{-4/3}\right) ^{1/2},\text{ }z\rightarrow\infty,
\label{eq24}
\end{equation}
respectively. In Fig. \ref{fig6} we plot $-\left\vert t\right\vert ^{-2/3}M/T$ \textit{vs}. $H_{c}\left\vert t\right\vert ^{-4/3}$ for $T<T_{\rm c}$ at various applied magnetic fields. The solid line is Eq. (\ref{eq23}) in terms of
\begin{eqnarray}
-\left\vert t\right\vert ^{-2/3}\frac{M}{T}&=&-3.32\times 10^{-6}\cdot \label{eq25}\\ 
&\textrm{ }&\left( \text{ln}(H_{c}\left\vert t\right\vert ^{-4/3})-11.23\right) \left( \text{emuK}^{-1}\right)
\nonumber
\end{eqnarray}
and the dashed one
\begin{equation}
-\left\vert t\right\vert ^{-2/3}\frac{M}{T}=2.3\times 10^{-9}(H_{c}\left\vert t\right\vert ^{-4/3})^{1/2}\text{ }\left( \text{emuK}^{-1}\right)
\label{eq26}
\end{equation}
corresponds to the limit (\ref{eq24}). While the $H_{c}\left\vert t\right\vert ^{-4/3}\propto z\rightarrow \infty $ limiting behavior (dashed curve) is well confirmed we observe with decreasing $z$ substantial deviations from the data collapse on a single curve.
%%%%%%%%%%%%%%%%

%%%%%%%%%%%%%%%%
\begin{figure}[t!]
\includegraphics[width=1\linewidth]{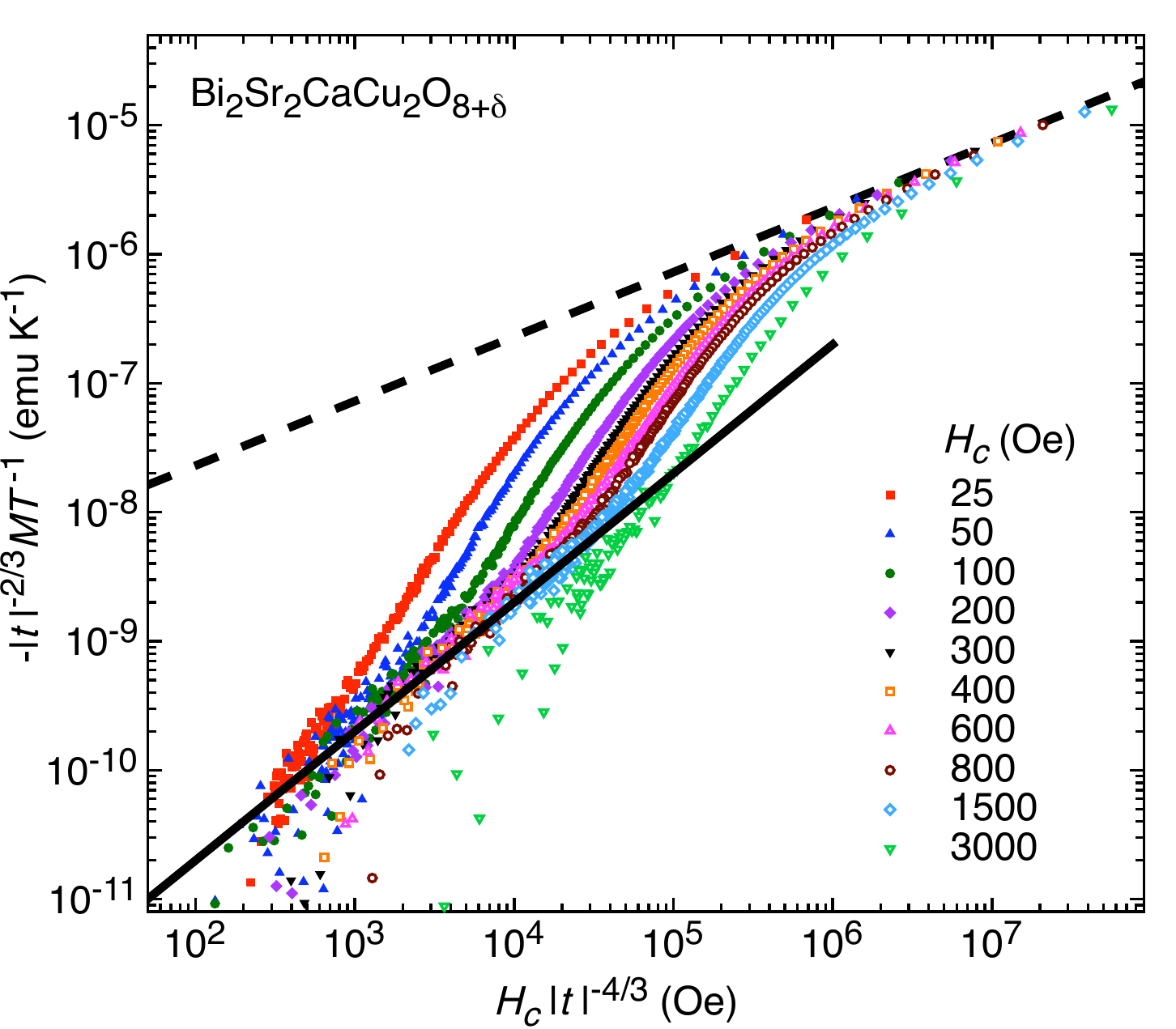}
\vspace{0cm}
\caption{(color online) $-\left\vert t\right\vert ^{-2/3}M/T$ \textit{vs}. $H_{c}\left\vert t\right\vert ^{-4/3}$ for $T>T_{\rm c}$ at various applied magnetic fields. The solid line is Eq. (\ref{eq30}) and the dashed one Eq. (\ref{eq31}).}
\label{fig7}
\end{figure}
%%%%%%%%%%%%%%%%

%%%%%%%%%%%%%%%%
\begin{figure}[b!]
\includegraphics[width=1\linewidth]{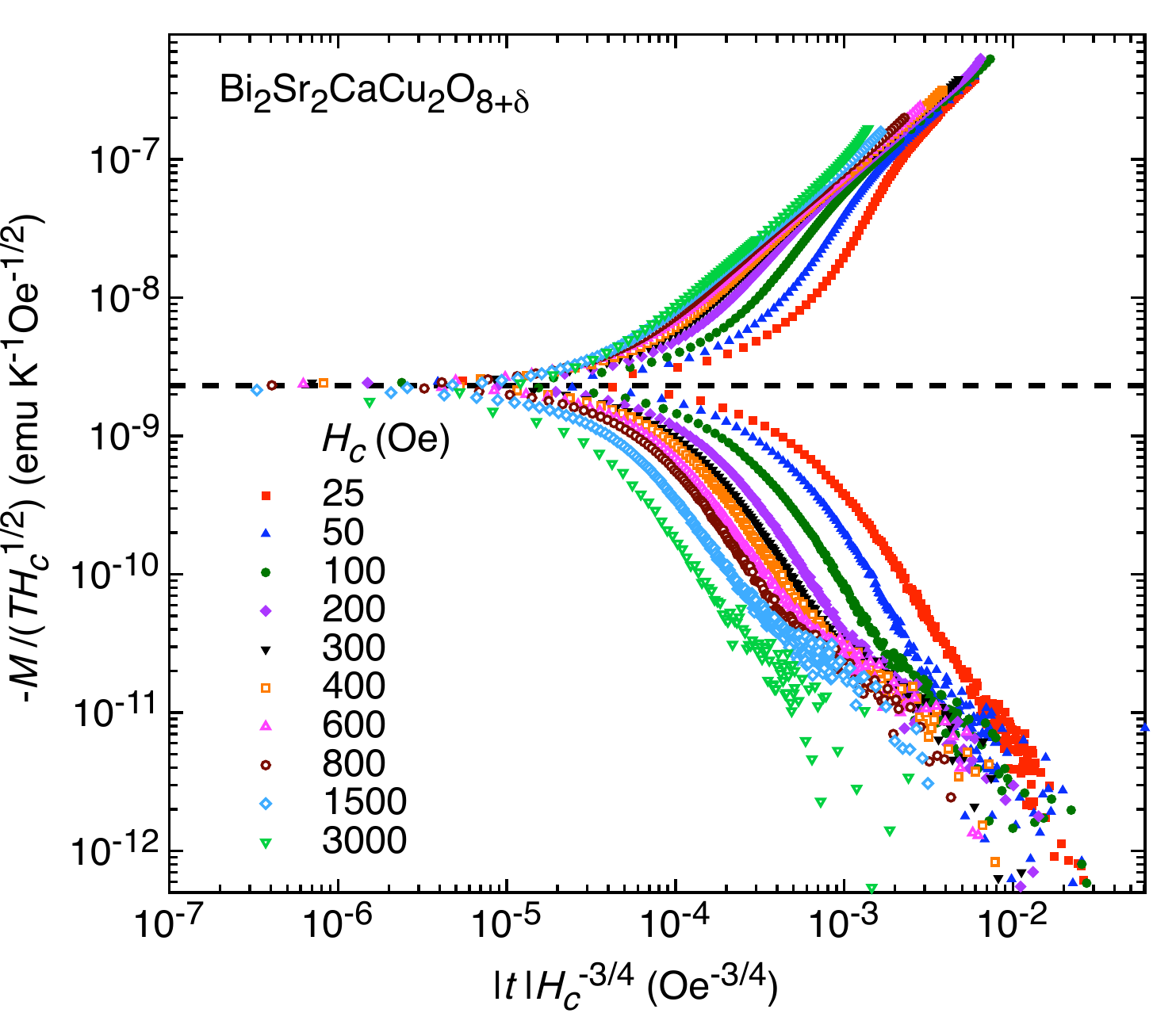}
\vspace{0cm}
\caption{(color online) $-M/(TH_{c}^{1/2})$ $vs.$ $|t|H_c^{-3/4}$ at various applied magnetic fields. The dashed line is Eq. (\ref{eq31}).}
\label{fig8}
\end{figure}
%%%%%%%%%%%%%%%%

From Eq. (\ref{eq26}) we derive with $V=4.6\times 10^{-5}$ cm$^{3}$, where $m=M/V$, the estimate
\begin{equation}
\gamma =\xi _{ab0}^{-}/\xi _{c0}^{-}\simeq 68,
\label{eq27}
\end{equation}
in reasonable agreement with $\gamma =\xi _{ab0}^{-}/\xi _{c0}^{-}\simeq 69$ (Eq. (\ref{eq19})), derived from the crossing point in $M/H_{c}^{1/2}$ \textit{vs}. $T$ shown in Fig. \ref{fig2}. The apparent failure of 3D-$xy$ scaling outside the limit $H_{c}\left\vert t\right\vert ^{-4/3}\propto z\rightarrow\infty $ is then attributable to the 3D- to 2D-$xy$ crossover. In particular, very small magnetic fields would be required to attain the limit $H_{c}\left\vert t\right\vert ^{-4/3}\propto z\rightarrow 0$ with reduced temperatures $t$ in the range where 3D-$xy$ fluctuations dominate. Indeed, given our evidence for 2D-$xy$ behavior above $T_{\rm KT}\simeq 89.5$ K in terms of the crossing phenomenon (Fig. \ref{fig3}) and 3D-$xy$ critical behavior above $T=91 $ K (Fig. \ref{fig4}b), transforming to $H_{c}\left\vert t\right\vert^{-4/3}=2\times 10^{6}$ Oe for $H_{c}=600$ Oe. Fig. \ref{fig6} reveals that below this value pronounced deviations from the expected data collapse occur, so in this regime 3D-$xy$ scaling fails because 3D-fluctuations no longer dominate. Nevertheless, at much lower fields the limit $H_{c}\left\vert t\right\vert^{-4/3}\propto z\rightarrow 0$ should be attainable, leading to the asymptotic behavior indicated by the solid line (Eq. (\ref{eq25})) in Fig. \ref{fig6}.
%%%%%%%%%%%%%%%%

Noting again that the occurrence of 3D behavior requires that the $c$-axis correlation length $\xi _{c}=\xi _{c0}^{\pm }\left\vert t\right\vert ^{-2/3}$ exceeds the interlayer spacing $d$ it is clear that the 3D- to 2D-$xy$ crossover is not restricted to temperatures below $T_{\rm c}$. Given the universal ratio $\xi _{c0}^{+}\simeq \xi _{c0}^{-}/2.21$ (Eq. (\ref{eq4})) it even follows that above $T_{\rm c}$ the 3D-$xy$ critical regime is even much narrower. Considering then the limits $H_{c}\left\vert t\right\vert^{-4/3}\propto z\rightarrow 0$ and $z\rightarrow \infty $ above $T_{\rm c}$, Eqs. (\ref{eq1}), (\ref{eq9}) and (\ref{eq10}) yield the scaling forms
\begin{equation}
\left\vert t\right\vert ^{-2/3}\frac{m}{T}=-\frac{Q^{+}c_{0}^{+}k_{\rm B}\left(\xi _{ab0}^{-}\right) ^{2}}{\Phi _{0}^{2}\xi _{c0}^{-}}H_{c}\left\vert t\right\vert ^{-4/3},z\rightarrow 0
\label{eq28}
\end{equation}
and
\begin{equation}
\left\vert t\right\vert ^{-2/3}\frac{m}{T}=-\frac{Q^{+}c_{\infty}^{+}k_{\rm B}\xi _{ab0}^{+}}{\Phi _{0}^{3/2}\xi _{c0}^{+}}\left(H_{c}\left\vert t\right\vert ^{-4/3}\right) ^{1/2},z\rightarrow \infty .
\label{eq29}
\end{equation}
In Fig. \ref{fig7} we depicted $-\left\vert t\right\vert^{-2/3}M/T$ \textit{vs}. $H_{c}\left\vert t\right\vert ^{-4/3}$ for $T>T_{\rm c}$ at various applied magnetic fields. The solid line is Eq. (\ref{eq28}) in terms of
\begin{equation}
-\left\vert t\right\vert ^{-2/3}M/T=2\times 10^{-13}H_{c}\left\vert t\right\vert ^{-4/3}\left( \text{emuK}^{-1}\right)
\label{eq30}
\end{equation}
and the dashed one
\begin{equation}
-\left\vert t\right\vert ^{-2/3}M/T=2.3\times 10^{-9}(H_{c}\left\vert t\right\vert ^{-4/3})^{1/2}\left( \text{emuK}^{-1}\right)
\label{eq31}
\end{equation}
corresponding to the limit (\ref{eq29}). Note that Eqs. (\ref{eq26}) and (\ref{eq31}) fully agree with each other and with that confirm $Q^{+}c_{\infty}^{+}=Q^{-}c_{\infty }^{-}$ (Eq. (\ref{eq10})). In analogy to Fig. \ref{fig6} we observe that the $z\rightarrow \infty $ limiting behavior is well confirmed, while substantial deviations from a data collapse on a single curve set in for $H_{c}\left\vert t\right\vert ^{-4/3}\lesssim 2\times 10^{6}$ Oe. Here the crossover to 2D-$xy$ behavior sets in and 3D-$xy$ scaling fails. Nevertheless, in the limit $H_{c}\rightarrow 0$ and reduced temperatures $t$ in the range where 3D-$xy$ fluctuations dominates, the limit $H_{c}\left\vert t\right\vert ^{-4/3}\propto z\rightarrow 0$ should be attainable. It is indicated by the solid line (Eq. (\ref{eq30})) in Fig. \ref{fig7}.
%%%%%%%%%%%%%%%%

According to Eq. (\ref{eq1}) one expects that the data plotted as $M/(TH_{c}^{1/2})$ $vs.$ $|t|H_c^{-3/4}$ should fall on two branches. An upper branch corresponding to $T>T_{\rm c}$ and a lower one for $T<T_{\rm c}$. A glance at Fig. \ref{fig8}, depicting this plot, clearly reveals the flow to 3D-$xy$ critical behavior by approaching $T_{\rm c}$ $(|t|=0)$ and in particular consistency with the leading 3D-$xy$ critical behavior below $|t|H_c^{-3/4}=10^{-5}$ Oe$^{-3/4}$.
%%%%%%%%%%%%%%%%

%%%%%%%%%%%%%%%%
\begin{figure}[t!]
\includegraphics[width=1\linewidth]{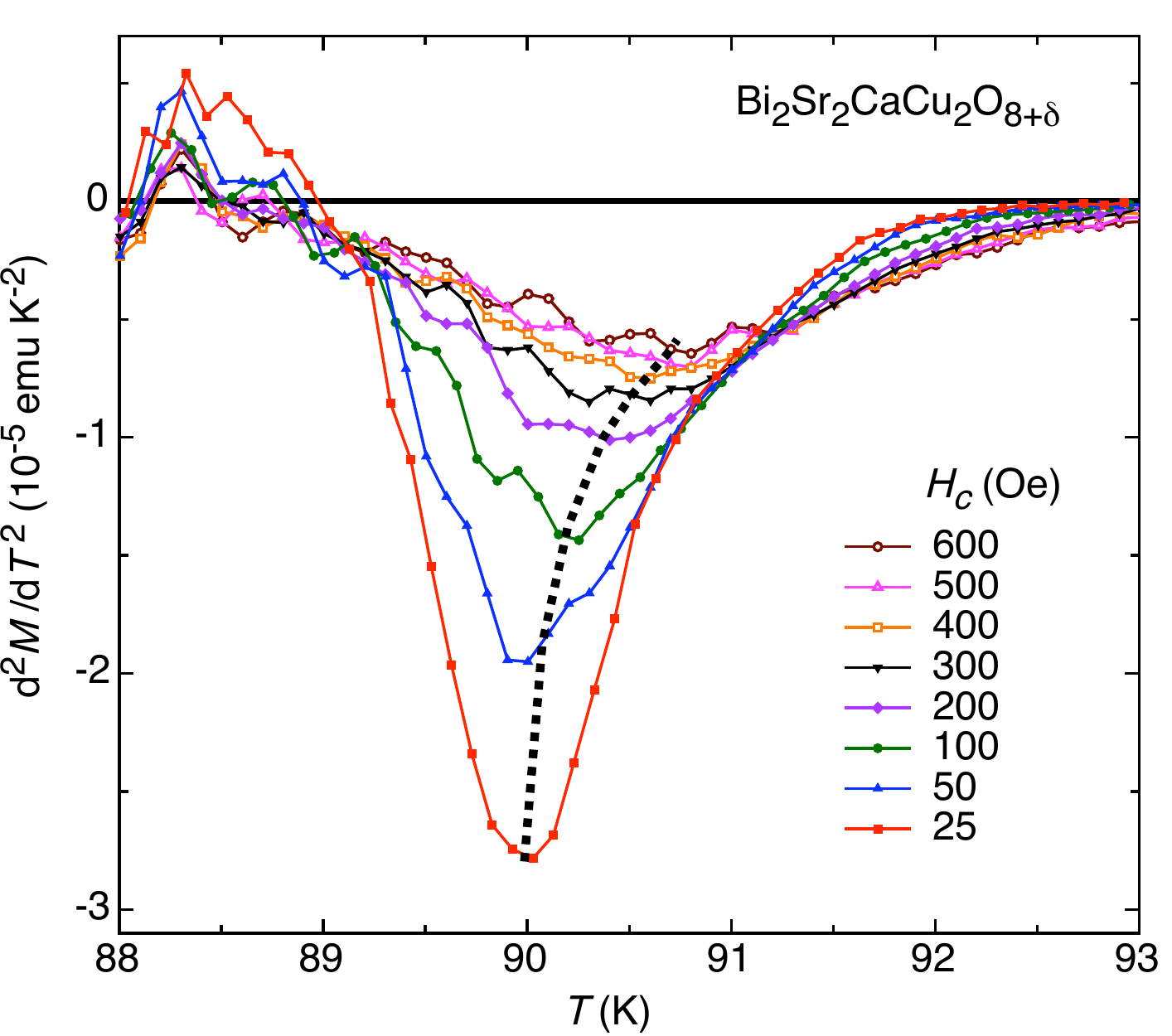}
\vspace{0cm}
\caption{(color online) $\left. \partial ^{2}M/\partial T^{2}\right\vert _{H_{c}}$ \textit{vs}. $T$\ at various $H_{c}$ according to Eq. (\ref{eq32}). The dotted line indicates the locations of the respective minima $T_{\rm p}(H_c)$.}
\label{fig9}
\end{figure}
%%%%%%%%%%%%%%%%

Another property where KT-behavior in terms of the 2D- to 3D-$xy$ crossover should be observable is the magnetic field dependence of the specific peak. Indeed, in Bi$_{2}$Sr$_{2}$CaCu$_{2}$O$_{8+\delta }$ this peak shifts,\cite{junod} opposite to the generic behavior.\cite{maxwts, Weyeneth2} Given the Maxwell relation,
\begin{equation}
\left. \frac{\partial ^{2}M}{\partial T^{2}}\right\vert _{H_{c}}=\left.\frac{\partial }{\partial H_{c}}\left( C/T\right) \right\vert _{T},
\label{eq32}
\end{equation}
relating magnetization and specific heat, \ this abnormality should also be observable in$\left. \partial ^{2}M/\partial ^{2}T\right\vert _{H_{c}}$. A glance at Fig. \ref{fig9}, depicting $\left. \partial ^{2}M/\partial T^{2}\right\vert_{H_{c}}$ \textit{vs}. $T$ for various magnetic fields reveals consistency with the anomalous shift of the specific heat peak. Indeed the dip shifts towards higher temperatures.
%%%%%%%%%%%%%%%%

\begin{figure}[t!]
\includegraphics[width=1\linewidth]{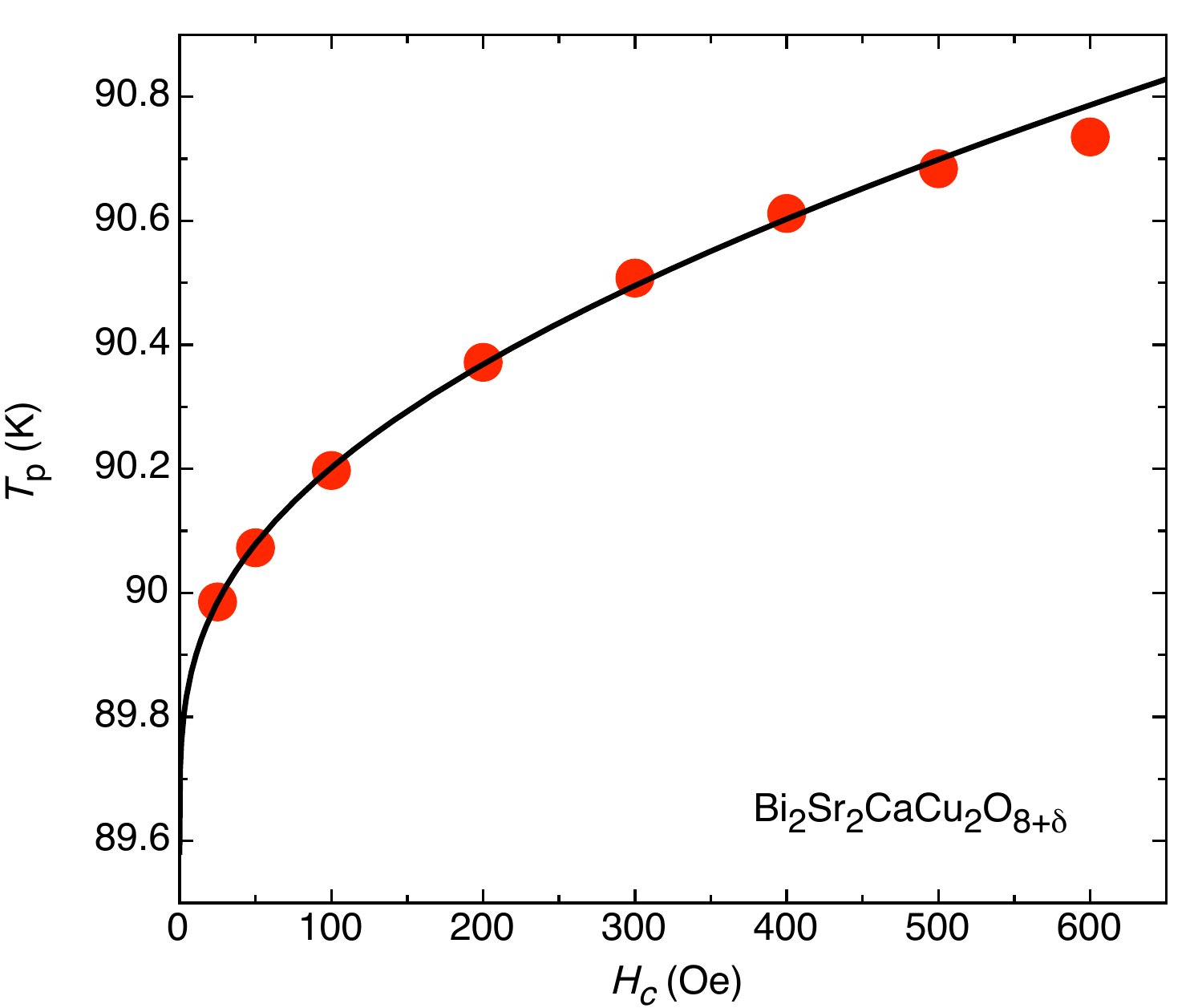}
\vspace{0cm}
\caption{(color online) $T_{\rm p}$ \textit{vs}. $H_{c}$ determined from the data depicted in Fig. \ref{fig9}. The solid line is Eq. (\ref{eq34}) with $T_{\rm KT}=89.5$ K, $\widetilde{b}\simeq0.3$, $a=3.12$, and $\xi _{ab0}=84$ \AA.}
\label{fig10}
\end{figure}

Noting that the anomalous shift occurs slightly above $T_{\rm KT}$ $\simeq 89.5$ K it is indeed expected to reflect 2D-$xy$ behavior. To check this conjecture we invoke expression (\ref{eq17}) for the magnetization valid above $T_{\rm KT}$ in the limit $H_{c}\rightarrow 0$. Although this limit is not attained in the field range considered here, $m$ is expected to scale as $m\simeq-\left( k_{\rm B}T/\left( 2d\Phi _{0}^{2}\right) \right) \xi _{ab}^{2}f\left(H_{c}\right) $ with $f\left( H_{c}\right) =H_{c}$ for $H_{c}\rightarrow 0$.\cite{tsepl} Accordingly, $\left. \partial ^{2}M/\partial T^{2}\right\vert_{H_{c}}$ diverges at $T_{\rm KT}$ \ However, there is the magnetic field induced finite size effect preventing the correlation length $\xi _{ab}$ to grow beyond the limiting magnetic length $L_{H_c}=\left( \Phi _{0}/\left(aH_c\right) \right) ^{1/2}$ where $a\simeq 3.12$.\cite{tsben,maxwts, Weyeneth1, Weyeneth2} As a consequence, in finite fields the divergence is removed by a dip. Its minimum occurs at $T_{\rm p}$ given by
\begin{equation}
\xi _{ab}\left( T_{\rm p}\right) =L_{H_{c}}=\left( \Phi _{0}/\left(aH_{c}\right) \right) ^{1/2}.
\label{eq33}
\end{equation}
In Fig. \ref{fig10} we show $T_{\rm p}$ \textit{vs}. $H$ derived from the data depicted in Fig. \ref{fig9}. For comparison we included
\begin{equation}
T_{\rm p}\left( H_{c}\right) =T_{\rm KT}\left( 1+\left( 2\widetilde{b}/\text{ln}(\frac{\Phi _{0}}{a\xi _{ab0}^{2}H_{c}})\right) ^{2}\right) ,  \label{eq34}
\end{equation}
resulting from Eqs. (\ref{eq17}) and (\ref{eq33}), for a realistic set of parameters. Apparently, the characteristic shift towards higher temperatures and the ln$(H_{c})$ behavior are well confirmed and the values for $\widetilde{b}$ (Eq. (\ref{eq17})) and $\widehat{b}$ (Eq. (\ref{eq14})) are reasonably consistent with the relation $\widetilde{b}\widehat{b}=\pi/\left( 2T_{\rm KT}^{1/2}\right) \simeq 0.17$ K$^{-1/2}$ (Eq.(\ref{eq18})). Furthermore, the observed $H_{c}$ dependence of $T_{\rm p}$ down to $25$ Oe also implies that the lateral extent $L_{ab}$ of the homogenous regions exceeds $L_{H_{c}}\simeq3640$ \AA . Indeed, in the opposite case ($L_{ab}<L_{H_{c}}$) $T_{\rm p}$ would be independent of $H_{c}$. Thus, $L_{ab}>3640$ \AA\ uncovers a remarkable sample homogeneity. Nevertheless it appears to be unlikely that the asymptotic $H_{c}\rightarrow 0$ is experimentally attainable. Here $T_{\rm p}\left( H_{c}\right) $ follows from $\xi _{ab}=\xi _{ab0}^{-}\left\vert T_{\rm p}\right\vert ^{-2/3}=L_{H_{c}}$ so
\begin{equation}
T_{\rm p}=T_{\rm c}\left( 1-\left( \frac{aH_{c}\left( \xi _{ab0}^{-}\right) ^{2}}{\Phi _{0}}\right) ^{3/4}\right),
\label{eq35}
\end{equation}
whereupon $T_{\rm p}$ shifts with increasing field to lower temperatures, as observed in a variety of less anisotropic cuprates.\cite{maxwts, Weyeneth2}
%%%%%%%%%%%%%%%%

%%%%%%%%%%%%%%%%
%%%%%%%%%%%%%%%%

%%%%%%%%%%%%%%%%
\section{Summary and discussion}
%%%%%%%%%%%%%%%%
Even though the mechanism of superconductivity in the cuprates remains mystery the associated phase transition properties can be understood as consequences of thermal fluctuations within the framework of the theory of critical phenomena. In this work we presented and analyzed reversible magnetization data of the highly anisotropic Bi$_{2}$Sr$_{2}$CaCu$_{2}$O$_{8+\delta }$ for magnetic fields applied along the $c$-axis of the high quality single crystal. We examined the occurrence of 3D-$xy$ critical behavior close to the bulk transition temperature $T_{\rm c}$ and of Kosterlitz-Thouless behavior. Below $T_{\rm c}$ and above the presumed Kosterlitz-Thouless transition temperature $T_{\rm KT}$ we observed, in agreement with the theoretical prediction,\cite{oganesyan} a downward shift of the \textquotedblleft crossing point\textquotedblright\ towards $T_{\rm KT}$ from above as the field is decreased. Sufficiently below $T_{\rm KT}$ we verified the characteristic 2D-$xy$ relationship between the magnetization an the in-plane magnetic penetration depth.\cite{oganesyan} In contrast, we have seen that the measured temperature dependence of the superfluid density does not exhibit the characteristic KT-behavior (Nelson-Kosterlitz jump) around the presumed $T_{\rm KT}$. The absence of this feature was traced back to the 2D- to 3D-$xy$ crossover setting in around and above $T_{\rm KT}$. Indeed, in the limit $H_{c}\left\vert t\right\vert^{-4/3}\rightarrow \infty $ we established clear evidence for 3D-$xy$ critical behavior above and below $T_{\rm c}$, while in the opposite limit ($H_{c}\left\vert t\right\vert ^{-4/3}\rightarrow 0$) its failure was attributed to the dimensional crossover. Invoking the Maxwell relation $\left. \partial ^{2}M/\partial ^{2}T\right\vert _{H}=\left. \partial \left(C/T\right) /\partial H_{c}\right\vert _{T}$ the anomalous field dependence of the specific heat peak was also traced back to the intermediate 2D-$xy$ behavior.\cite{junod} Implications include: First, sufficiently below $T_{\rm KT}$ \ the isotope and pressure effects on $dm/d\ln (H_{c})$ at fixed temperature $T$ and the in-plane magnetic penetration depth $\lambda_{ab}\left( T\right) $ are not independent but related by Eq. (\ref{eq15}). Second, the bulk transition temperature $T_{\rm c}$ and the critical amplitudes of the $c$-axis correlation length and in-plane magnetic penetration depth are related by the universal relation (\ref{eq8}), so the isotope and pressure effects on these properties are related.\cite{tsiso}
%%%%%%%%%%%%%%%%

\section{Acknowledgments}
This work was partially supported by the Swiss National Science Foundation and the EU Project CoMePhS.

\end{document}